\documentclass[english,reprint,pre,amsmath]{revtex4-1}
\usepackage[T1]{fontenc}
\usepackage[latin9]{inputenc}
\usepackage{array}
\usepackage{mathrsfs}
\usepackage{amsmath}
\usepackage{amsthm}
\usepackage{amssymb}
\usepackage{stmaryrd}
\usepackage{graphicx}

\makeatletter

\usepackage{color}  
\RequirePackage[dvipsnames,usenames]{xcolor}
\usepackage[colorlinks,linkcolor=blue,citecolor=blue]{hyperref}
\usepackage{newtxtext}

\AtBeginDocument{\providecommand\figref[1]{Fig.~\ref{fig:#1}}}
\AtBeginDocument{\providecommand\secref[1]{Sec.~\ref{sec:#1}}}
\AtBeginDocument{\providecommand\tabref[1]{Table~\ref{tab:#1}}}
\providecommand{\tabularnewline}{\\}
\newcommand{\newstuff}[1] { #1 }

\makeatother

\usepackage{babel}
\begin{document}
\global\long\def\mysection#1{\vspace{2pt}\emph{#1}.--- }%
\global\long\def\xs{x}%
\global\long\def\xf{y}%
\global\long\def\hdbits{10^{11}}%
\global\long\def\KK{K(\xs\vert\xf)}%
\global\long\def\utm{\mathfrak{C}}%
\global\long\def\uconst{\gamma_{\utm,\PP}}%
\global\long\def\uuconst{\gamma_{\utm}}%
\global\long\def\BB{\{0,1\}^{*}}%
\global\long\def\xM{x\shortrightarrow y}%
\global\long\def\PP{\mathcal{P}}%
\global\long\def\nrg{\epsilon}%
\global\long\def\tr{\mathrm{tr}}%
\global\long\def\OO{\mathcal{O}}%
\global\long\def\bs{b}%
\global\long\def\bf{c}%
\global\long\def\ctraj{H}%
\global\long\def\miK{I^{K}}%

\global\long\def\nrg{\epsilon}%
\global\long\def\basisX{\mathcal{A}}%
\global\long\def\bathi{b}%
\global\long\def\bathf{c}%
\global\long\def\soiX{A}%

\global\long\def\bath{B}%
\global\long\def\basisE{\mathcal{B}}%
\global\long\def\bathx{b}%
\global\long\def\sysstate{a}%

\global\long\def\eEnt{\mathscr{E}}%

\global\long\def\inputP{p(x)}%

\global\long\def\lntb{{\textstyle \frac{\ln2}{\beta}}}%
\global\long\def\lntbIn{\lntb}%
\global\long\def\blnt{{\textstyle \frac{\beta}{\ln2}}}%
\global\long\def\blntIn{\blnt}%
\global\long\def\qqBase{Q(\xM)}%
\global\long\def\qq{\blnt\qqBase}%
\global\long\def\ppComp{K(\mathcal{P})}%
\global\long\def\kk{\mathcal{I}}%
\global\long\def\kkComp{\mathcal{I}(\PP\!:x\vert y)}%
\global\long\def\KKp{K(\xs\vert\xf,\PP^{*})}%
\global\long\def\revP{\tilde{p}}%

\title{Generalized Zurek's bound on the cost of an individual classical or
quantum computation}
\author{Artemy Kolchinsky}
\affiliation{Universal Biology Institute, The University of Tokyo, 7-3-1 Hongo,
Bunkyo-ku, Tokyo 113-0033, Japan}
\begin{abstract}
We consider the minimal thermodynamic cost of an individual computation,
where a single input $x$ is mapped to a single output $y$. In
prior work, Zurek proposed that this cost was given by $K(x\vert y)$,
the conditional Kolmogorov complexity of $x$ given $y$ (up to an
additive constant which does not depend on $x$ or $y$). However,
this result was derived from an informal argument, applied only to
deterministic computations, and had an arbitrary dependence on the
choice of protocol (via the additive constant). Here we use stochastic
thermodynamics to derive a generalized version of Zurek's bound from
a rigorous Hamiltonian formulation. Our bound applies to all quantum
and classical processes, whether noisy or deterministic, and it explicitly
captures the dependence on the protocol. We show that $K(x\vert y)$
is a minimal cost of mapping $x$ to $y$ that must be paid using
some combination of heat, noise, and protocol complexity, implying
a tradeoff between these three resources. Our result is a kind of
``algorithmic fluctuation theorem'' with implications for the relationship
between the Second Law and the Physical Church-Turing thesis.
\end{abstract}
\maketitle

\section{Introduction}

It is now understood that there are fundamental relationships between
computational and thermodynamic properties of physical processes.
The best-known relationship is Landauer's bound, which says that any
computational process that erases statistical information must generate
a corresponding amount of thermodynamic entropy in its environment
\citep{landauer1961irreversibility,benn82}. For concreteness, imagine
a process that implements some stochastic input-output map $p(y\vert x)$
while coupled to a heat bath at inverse temperature $\beta$. Suppose
that the process is initialized with some ensemble of inputs $\inputP$
which is mapped to an ensemble of outputs $p(y)=\sum_{x}p(y\vert x)p(x)$.
Landauer's bound implies that the generated heat, averaged across
the ensemble of system trajectories, obeys
\begin{equation}
\blnt\langle Q\rangle\ge S(p(X))-S(p(Y)),\label{eq:landauer0}
\end{equation}
where $S(\cdot)$ is the Shannon entropy in bits. The result imposes
a minimal ``thermodynamic cost of computation'', i.e., a minimal
amount of internal energy and/or work that must be lost as heat.

Importantly, Landauer's bound depends not only on properties of the
physical process but also on the choice of the input ensemble $\inputP$.
Because of this, it cannot be used to investigate the following natural
question: what is the cost of mapping a single input $\xs$ to a
single output $\xf$, independently of which statistical ensembles
(if any) the inputs are drawn from? As a motivating example, imagine
a process that deterministically maps the logical state of a 100GB
hard drive from some particular sequence of $10^{11}$ zeros-and-ones
to a sequence of $10^{11}$ zeros. Without additional assumptions
about the input ensemble, one cannot use Landauer's bound to constrain
the heat generated by this process. The same holds for most other
bounds on the thermodynamic costs of computation, which typically
depend on the choice of the input ensemble  \citep{maroney2009generalizing,faist2015minimal,parrondo2015thermodynamics,kolchinsky2016dependence,boyd2016identifying,ouldridge2017fundamental,Boyd:2018aa,wolpert2019stochastic,wolpert2020thermodynamic,riechers2021initial,riechersImpossibilityLandauerBound2021,kolchinsky2021dependence,kardecs2022inclusive}.

\begin{figure}[b]

\includegraphics[width=1\columnwidth]{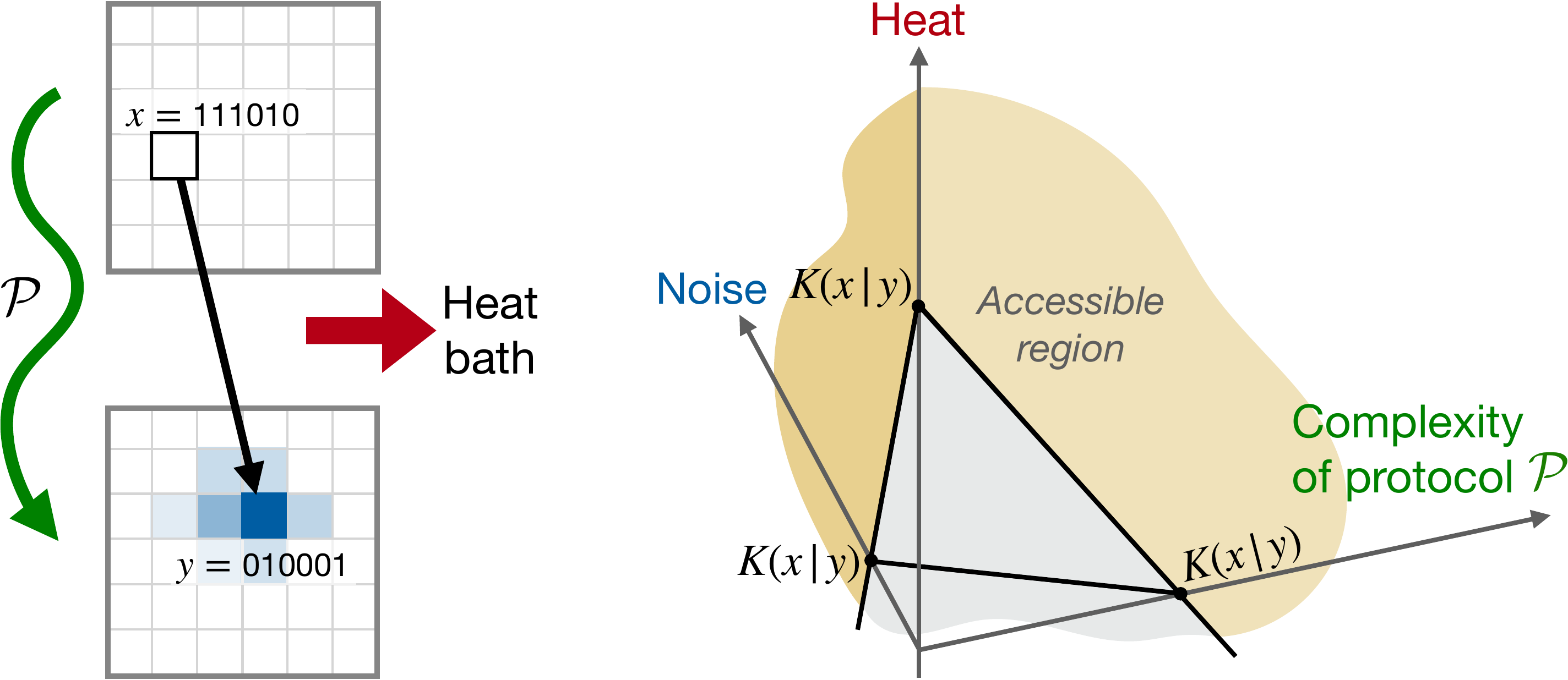}

\caption{\label{fig:1}We analyze the cost of mapping a single input $x$ to
a single output $y$, given a system coupled to a heat bath and a
work reservoir that drives protocol $\mathcal{P}$. We show that the
conditional Kolmogorov complexity $K(x\vert y)$ is a fundamental
minimal cost, which bounds accessible combinations of heat, noise,
and complexity of the protocol.}
\end{figure}

In this paper, we derive a lower bound on the heat generated by an
individual computation that maps a single input $x$ to a single
output $y$ (see \figref{1}). Our bound, which appears in Section~\ref{sec:bound}
below, depends only on the properties of $x$, $y$, and the protocol
that carries out the transformation $\xM$. It reflects the loss of
\emph{algorithmic information} in going from $x$ to $y$ \citep{livi08},
rather than the loss of \emph{statistical information} in going from
the ensemble $\inputP$ to $p(y)$, as in Landauer's bound. 

To derive our result, we suppose without loss of generality that $x$
and $y$ can be represented by binary strings. The loss of algorithmic
information is quantified as the length of the shortest computer program
that outputs $x$ when provided with $y$ as input:
\begin{equation}
K(x\vert y)=\min_{s}\ell(s)\quad\text{such that}\quad\utm(s,y)=x.\label{eq:kdef}
\end{equation}
Here the minimization is over all binary strings, $\ell(s)$ indicates
the length of string $s$, and $\utm(s,y)$ indicates the output of
a fixed universal computer $\utm$ on program $s$ and input $y$.
In algorithmic information theory, $K(x\vert y)$ is termed the conditional
Kolmogorov complexity of $x$ given $y$ \citep{livi08,vitanyi2013conditional}. 

\newstuff{The conditional Kolmogorov complexity quantifies how much
information must be added to $y$ in order to recover $x$, and so
it can be understood as the amount of algorithmic information about
$x$ that is not in $y$.} It is the algorithmic analog of conditional
entropy in Shannon's information theory, and it has many remarkable
mathematical properties. Since it is defined at the level of individual
strings, rather than statistical ensembles like Shannon entropy, it
is well-suited for studying the cost of individual computations.

Our approach builds on the work by Zurek from the late 1980s \citep{zure89a},
which suggested that $K(x\vert y)$ can be used to bound the cost
of an individual computation $\xM$ without reference to any statistical
ensembles (see also \citep{zure89b,bennett1993thermodynamics,bennett1998information,baumeler2019free}).
Zurek considered a system in contact with an environment, such as
a heat bath at inverse temperature $\beta$, which maps some input
$x$ to output $y$ in a deterministic (noiseless) manner. He argued
that the heat generated by the individual computation, which we write
as $\qqBase$, obeys
\begin{equation}
\qq\ge K(x\vert y)+O(1),\label{eq:zurek0}
\end{equation}
where $O(1)$ refers to an arbitrary constant that does not depend
on $x$ or $y$, but does depend on the choice of the universal computer
$\utm$ and the protocol that implements the computation. \newstuff{Zurek's
bound may be compared to Landauer's bound (\ref{eq:landauer0}), since
the latter can be written as ${\textstyle \blntIn}\langle Q\rangle\ge S(p(X\vert Y))$
in the case where $X$ is a deterministic function of $Y$. The conditional
Shannon entropy $S(p(X\vert Y))$ quantifies the loss of statistical
information about the input given the output, averaged across the
ensemble of computational trajectories; the conditional Kolmogorov
complexity $K(x\vert y)$ quantifies the loss of algorithmic information
about the input given the output at the level of a single computation.
Importantly, Zurek's bound is not simply a special case of Landauer's
bound where the initial ensemble is concentrated on a single input
$x$, since in that case $S(p(X\vert Y))$ vanishes while $K(x\vert y)$
generally does not.}

Zurek's bound suggests a remarkable connection between thermodynamics
and algorithmic information theory. However, it has some drawbacks.
First, it is limited to deterministic computations. Second, it is
only meaningful when considering the behavior of a fixed protocol
on different inputs. In fact, for any particular $x$ and $y$, one
can construct a special protocol that computes $\xM$ while generating
an arbitrarily small amount of heat \citep[Sec. V, ][]{kolchinsky2020thermodynamic}.
Finally, as we discuss below, Zurek's bound was derived using an informal
argument that invoked Landauer's bound as an intermediate step. This
is problematic because, as mentioned above, Landauer's bound only
applies at the level of statistical ensembles, not individual computations.
This suggests that Zurek's original derivation made an implicit assumption
about the statistical distribution of inputs.

Our result is a generalized version of Zurek's bound that overcomes
all of these issues. It holds for classical and quantum processes,
noisy as well as deterministic. In addition, it explicitly quantifies
how the choice of protocol enters into the bound, via a ``protocol
complexity'' term. Finally, our bound is derived in an explicit manner
within a rigorous Hamiltonian framework. Our derivation uses a combination
of techniques from algorithmic information theory and stochastic thermodynamics,
which together allow us to study the thermodynamic properties of individual
computational trajectories. In general, our work contributes to the
growing body of recent research on the relationship between algorithmic
information theory and nonequilibrium thermodynamics \citep{hiura2019microscopic,kolchinsky2020thermodynamic,ro2022model,wolpert2019stochastic,baumeler2019free}.

Although our result is motivated within the context of the thermodynamics
of computation, it is not restricted to systems that are conventionally
considered to be computers. In fact, it is a general algorithmic bound
on the heat generated by any system that undergoes the transformation
$\xM$ while coupled to a thermal reservoir. For this reason, it may
also be used to study the thermodynamic properties of individual instances
of other types of processes, including work extraction \citep{rio2011thermodynamic,skrzypczyk2014work,hovhannisyan2020charging}
and algorithmic cooling \citep{schulman2005physical,allahverdyan2011thermodynamic,raeisi2015asymptotic}
protocols.

Our paper is structured as follows. In the next section, we provide
preliminaries and discuss our setup. We derive our main results in
\secref{bound} and illustrate it with examples in \secref{Example}.
We relate our approach to prior work in \secref{Relation-to-prior}.
A brief discussion follows in \secref{conclusion}. Derivations of
our bound and its achievability are found in the appendices.

\section{Preliminaries}

\label{sec:Setup}

\subsection{Physical setup}

We consider a simple and operationally accessible physical setup,
inspired by ``two-point measurement'' schemes used in quantum thermodynamics
\citep{tasaki2000jarzynski,esposito2009nonequilibrium,campisi2011colloquium,manzanoQuantumFluctuationTheorems2018,BuscemiBayesianRetrodiction2021}.
We suppose that there is a \emph{computational subsystem} $\soiX$ which
carries out the transformation $\xM$. This subsystem is coupled to
a thermal bath, represented by subsystem $\bath$. The overall setup
may be classical or quantum, though here we focus on the quantum case
for generality.

We assume that the Hilbert spaces of $\soiX$ and $\bath$ are both
separable, so that each one can be associated with a countable orthonormal
basis. The basis of subsystem $\soiX$ is indexed by some set of binary
strings $\basisX\subseteq\{0,1\}^{*}$, which we indicate as $\{\vert\sysstate\rangle\}_{\sysstate\in\basisX}$.
(The notation $\{0,1\}^{*}$ refers to the countable set of binary
strings of finite, but arbitrary, length.) The basis vectors $\vert\sysstate\rangle$
represent logical states, so that the computation $\xM$ corresponds
to a physical process during which subsystem $\soiX$ goes from
the initial state $\vert x\rangle$ to the final state $\vert y\rangle$.
The environment $\bath$ is a heat bath at inverse temperature $\beta$
and Hamiltonian $H_{\bath}$. We indicate the spectral decomposition
of the Hamiltonian as $H_{\bath}=\sum_{\bathx\in\mathcal{\basisE}}\nrg_{\bathx}\vert\bathx\rangle\langle\bathx\vert,$
where $\basisE$ indexes the bath's energy eigenstates as $\{\vert\bathx\rangle\}_{\bathx\in\basisE}$.
We indicate the product basis formed by the logical states of $\soiX$
and the energy eigenstates of $\bath$ as $V=\{\vert\sysstate,\bathx\rangle\}_{\sysstate\in\basisX,\bathx\in\basisE}$.

The computation $\xM$ is carried out as follows. $\soiX$ and $\bath$
start in a pure product state $\vert x,\bathi\rangle\in V$, where
$\vert\bathi\rangle$ is sampled from the Gibbs distribution $\pi(\bathi)=e^{-\beta\nrg_{\bathi}}/Z$.
The two subsystems are then coupled to an external work reservoir
and undergo a Hamiltonian driving protocol over time $t\in[0,\tau]$,
corresponding to a time-dependent Hamiltonian $H_{A\bath}(t)$. As
a result of this driving protocol, $\soiX$ and $\bath$ jointly evolve
according to the final state $U\vert x,\bathi\rangle$ under the unitary
$U=\mathcal{T}e^{-(i/\hbar)\int_{0}^{\tau}H_{A\bath}(t)\,dt}$, where
$\mathcal{T}$ is the time-ordered exponential. Finally, $\soiX$
and $\bath$ undergo a projective measurement in the product basis
$V$. Given initial state $\vert x,\bathi\rangle$, the probability
of measuring output logical state $\vert y\rangle$ and bath energy
eigenstate $\vert\bathf\rangle$ is 
\begin{equation}
p(y,\bathf\vert x,\bathi)=\vert\langle y,\bathf\vert U\vert x,\bathi\rangle\vert^{2}.\label{eq:cond0}
\end{equation}
We refer to the combination of the product basis $V$, the unitary
$U$, the inverse temperature $\beta$, and the bath's energy function
$\epsilon$ as ``the protocol $\PP=(V,U,\beta,\epsilon)$''.

We will be interested in two properties of the computation $\xM$
as instantiated by the protocol $\PP$. The first property is the
conditional probability of output $y$ given input $x$, averaged
across bath states:
\begin{equation}
p(y\vert x)=\sum_{\bathi,\bathf}p(y,\bathf\vert x,\bathi)\pi(\bathi).\label{eq:cond1}
\end{equation}
The second property is the heat generated by the computation $\xM$:
\begin{align}
\qqBase & =\sum_{\bathi,\bathf}\frac{p(y,\bathf\vert x,\bathi)\pi(\bathi)}{p(y\vert x)}(\nrg_{\bathf}-\nrg_{\bathi}).\label{eq:heat-def}
\end{align}
This is the average energy increase of the bath, conditioned on input
logical state $x$ and output logical state $y$.

\subsection{Computability and description of the protocol}

\label{subsec:Computability-assumption}

In order to derive our results, we make an important \emph{computability
assumption} in regards to the protocol $\PP=(V,U,\beta,\epsilon)$:
we assume that there exists a program for a universal computer $\utm$
that can approximate, to any desired degree of numerical precision,
the values of
\begin{equation}
\begin{aligned}\beta,\quad\nrg_{\bathi},\quad p(y,\bathf\vert x,\bathi) & =\vert\langle y,\bathf\vert U\vert x,\bathi\rangle\vert^{2} & \text{for all \ensuremath{x,b,y,c}.}\end{aligned}
\label{eq:PPprog}
\end{equation}
We discuss the physical meaning of the computability assumption at
the end of this paper.

\newstuff{We refer to the shortest program that computes the values
(\ref{eq:PPprog}) as $\PP^{*}$. We emphasize that any program that
can compute these values can also be used to compute the bath Gibbs
distribution $\pi(\bathi)=e^{-\beta\nrg_{\bathi}}/Z$, as well as
the entries of the stochastic input-output map $p(y\vert x)$
and the generated heat $\qqBase$, via (\ref{eq:cond1}) and (\ref{eq:heat-def})
respectively. Therefore, $\PP^{*}$ can be interpreted as the minimal
description of the relevant computational and thermodynamic properties
of the physical protocol. We refer to the length of this minimal description
$\PP^{*}$ as ``the complexity of protocol $\PP$'', and indicate
it as $\ppComp\equiv\ell(\PP^{*})$.

In deriving our results, we will make use of the conditional Kolmogorov
complexity $\KK$, as defined in (\ref{eq:kdef}). We also make use
of the conditional Kolmogorov complexity $K(x\vert y,\PP^{*})$, which
is defined in a similar way as the length of the shortest program
that outputs $x$ when provided with $y$ and the minimal description
$\PP^{*}$ as input. $K(x\vert y,\PP^{*})$ quantifies the algorithmic
information about $x$ that is not found in the combination of $y$
and $\PP^{*}$. It can be related to $\KK$ and $\ppComp$ via the
inequalities
\begin{equation}
\begin{aligned}K(x\vert y) & \ge\KKp+O(1)\\
K(x\vert y) & \le K(\PP)+\KKp+O(1),
\end{aligned}
\label{eq:df3}
\end{equation}
where $O(1)$ refers to additive constants that do not depend on $x$,
$y$, or $\PP$. The first inequality follows because additional side
information cannot increase conditional Kolmogorov complexity. The
second inequality follows because $K(x\vert y)$, the length of the
shortest program that outputs $x$ when provided with $y$, cannot
be longer than the length of $\PP^{*}$ plus a program that outputs
$x$ when provided with $y$ and $\PP^{*}$. Combining these inequalities
implies
\begin{equation}
\vert K(x\vert y)-\KKp\vert\le K(\PP)+O(1).\label{eq:simp1}
\end{equation}
}

We finish by noting a few technical points related to our use of algorithmic
information theory (AIT). As standard in AIT, we assume that the universal
computer $\utm$, which is used to define our Kolmogorov complexity
terms $K(x\vert y)$, $\KKp$, and $\ppComp$, accepts self-delimiting
programs. This means that the set of valid programs for $\utm$ forms
a prefix code \citep{cover_elements_2006,livi08}. We also note that
our Kolmogorov complexity terms depend on the choice of the universal
computer $\utm$, although we leave this dependence implicit in our
notation. A classic result in algorithmic information theory (AIT)
states that the choice of the universal computer only affects $K(x\vert y)$
by an additive constant: if $K(x\vert y)$ and $K'(x\vert y)$ are
defined using two different universal computers $\utm$ and $\utm'$,
then $\vert K(x\vert y)-K'(x\vert y)\vert\le O(1)$, where $O(1)$
refers to a constant that does not depend on $x$ or $y$ (only on
$\utm$ and $\utm'$). The same kind of invariance up to an additive
constant holds for $\ppComp$ and $\KKp$. A standard textbook reference
on AIT is Ref.~\citep{vitanyi2013conditional}. More succinct and
physics-oriented introductions, which are sufficient to understand
the content of this paper, can be found in Ref.~\citep{zure89a}
and Ref.~\citep{kolchinsky2020thermodynamic}.

\subsection{Use of the quantum formalism}

\label{subsec:quantum}

\newstuff{In this paper, we work in the quantum setting due to its
generality and preciseness, since any classical description is ultimately
an approximation of an underlying quantum physics. It is also convenient,
because the quantum formalism naturally leads to a countable basis
for the state space, which provides a discrete set of logical states
that can be studied using algorithmic information theory. In principle,
however, similar results can be derived for a classical Hamiltonian
system with a continuous phase space, as long as one introduces an
appropriate coarse-graining of the phase space into discrete logical
states \citep{deffner2013information}. A coarse-grained version of
our result can also be derived for quantum systems where the logical
states correspond to macrostates, rather than pure states. For simplicity,
we do not consider coarse-graining in this paper.

It is important to note that the stochastic map from initial to final
states of the computational subsystem and bath, $p(y,\bathf\vert x,\bathi)$
in (\ref{eq:cond0}), is a classical conditional probability distribution.
The stochastic input-output map over logical states, $p(y\vert x)$
in (\ref{eq:cond1}), is also a classical conditional probability
distribution. These conditional distributions are classical because
of the two-point measurement scheme considered here, in which the
system is initialized with classical information (the choice of a
pure state from a fixed reference basis) and outputs classical information
(the result of a projective measurement in a fixed reference basis).
At the same time, this does not preclude intermediate stages of the
computational process from exploiting quantum effects such as coherence
and entanglement. This setup is consistent with the process considered
by Zurek \citep{zure89a}, as well as the standard picture of quantum
computation in which an intermediate quantum process is used to map
classical inputs to classical outputs \citep{divincenzo1995quantum}.
However, as we touch upon in the Discussion, future work may extend
our analysis to a purely quantum formulation which does not require
fixed reference bases and projective measurements.

Because we consider the overall computation as a classical input-output
distribution, our results use standard algorithmic information theory,
as defined in terms of classical Turing machines, rather than one
of its quantum extensions \citep{vitanyi2001quantum,berthiaume2001quantum,mora2007quantum}.
We emphasize that there is no difficulty in describing a quantum protocol
$\PP$ using a classical Turing machine. This is because quantum states
and quantum operations can always be represented and manipulated on a classical
computer, for instance by using complex-valued matrices.}

\section{Main results}

\label{sec:bound}

\subsection{Algorithmic cost of an individual computation}

\label{subsec:Algorithmic-costs-of}

\newstuff{Our first main result is the following algorithmic bound
on any physically instantiated  computation $\xM$:
\begin{equation}
{\textstyle \blnt}\qqBase+\log_{2}\frac{1}{p(y\vert x)}+\uuconst\ge\KKp.\label{eq:resInt2}
\end{equation}
Here $\qqBase$ is the generated heat by the protocol $\PP$ that
carries out the computation, $-\log_{2}p(y\vert x)$ is the amount
of statistical noise, and $\KKp$ is the conditional Kolmogorov complexity
of input $x$ given output $y$ and the minimal description $\PP^{*}$,
as discussed above. Finally, $\uuconst$ is an additive constant which
depends on the universal computer $\utm$, but not $x$, $y$, or
$\PP$. This result implies that $\KKp$ is a fundamental cost of
carrying out the computation $\xM$ with protocol $\PP$, which must
be paid for either by heat or noise. A simple rearrangement of (\ref{eq:resInt2})
gives a lower bound on heat generation:
\begin{equation}
{\textstyle \blnt}\qqBase\ge\KKp-\log_{2}\frac{1}{p(y\vert x)}-\uuconst.\label{eq:resInt}
\end{equation}
Note that the right hand side of (\ref{eq:resInt}) may be negative,
in which case our result bounds the maximum heat that may be absorbed
from the heat bath.}

Our result does not depend on the choice of the input ensemble $\inputP$.
However, the noise term $-\log_{2}p(y\vert x)$ does depend on the
conditional output ensemble $p(y\vert x)$, which we treat as an intrinsic
property of the protocol. The noise term can be further decomposed
into separate classical and quantum contributions as
\begin{multline*}
\log_{2}\frac{1}{p(y\vert x)}=\\
\big\langle\mathscr{E}_{b}^{x}\big\rangle_{\pi}+\big\langle\log_{2}\frac{1}{p(y\vert x,b)}-\mathscr{E}_{b}^{x}\big\rangle_{\pi}+\big\langle\log_{2}\frac{p(y\vert x,b)}{p(y\vert x)}\big\rangle_{\pi},
\end{multline*}
where $\big\langle f_{\bathi}\big\rangle_{\pi}=\sum_{\bathi}\pi(\bathi)f_{\bathi}$
indicates expectations under the Gibbs distribution of the bath. In
this decomposition, $\eEnt_{b}^{x}$ is the von Neumann entropy of
the reduced state $\rho_{\soiX}^{x\bathi}=\tr_{\bath}\{U\vert x,\bathi\rangle\langle x,\bathi\vert U^{\dagger}\}$,
which is called ``entanglement entropy'' \citep{bennett1996concentrating}.
It is a nonnegative contribution which arises from quantum correlations
(entanglement) between the computational subsystem $\soiX$ and the
heat bath $\bath$. The second term $-\log_{2}p(y\vert x,\bathi)-\eEnt_{b}^{x}$
reflects the noise that arises from quantum coherence of the output
logical state in the measurement basis.  Averaged across initial
bath states $\pi(\bathi)$ and output logical states $p(y\vert x)$,
it equals $\big\langle S(Y\vert x,\bathi)-\eEnt_{b}^{x}\big\rangle\ge0$,
which is the expected ``relative entropy of coherence'' \citep{baumgratz_quantifying_2014}.
Finally, $\log_{2}[p(y\vert x,\bathi)/p(y\vert x)]$ arises due to statistical
uncertainty about the initial bath state. Averaged across initial
bath states $\pi(\bathi)$ and output logical states $p(y\vert x)$,
it equals $I(Y;B\vert x)\ge0$, the mutual information between initial
bath states $B$ and outputs $Y$ given input $x$. This is the contribution
from classical correlations between the $\soiX$ and $\bath$.

The derivation of (\ref{eq:resInt2}) is found in Appendix~\ref{app:derivation}.
This derivation is based on a rigorous Hamiltonian formulation and
does not impose any idealized assumptions on the heat bath, such as
infinite heat capacity, separation of time scales, etc. Moreover,
in Appendix~\ref{app:Achievability}, we show that the bound becomes
achievable, as long as the heat bath is nearly ideal. \newstuff{Specifically,
we imagine there is some desired computation $p(y\vert x)$, as well
as a nearly-ideal heat bath with inverse temperature $\beta$
and energy function $\epsilon$. Then, as long as $p(y\vert x)$,
$\beta$, and $\epsilon$ are computable, we demonstrate that there
are computable protocols that come close to equality in (\ref{eq:resInt2}).}
 
\newstuff{Our result is related to the so-called ``detailed fluctuation
theorem'' (DFT) in stochastic thermodynamics \citep{jarzynski2000hamiltonian,sagawa2012fluctuation,kwonFluctuationTheoremsQuantum2019,BuscemiBayesianRetrodiction2021}.
 In the particular setup considered here, the DFT can be used to derive
the bound
\begin{equation}
{\textstyle \frac{\beta}{\ln2}}\qqBase\ge\log_{2}\frac{1}{\tilde{p}(x\vert y)}-\log_{2}\frac{1}{p(y\vert x)},\label{eq:dft}
\end{equation}
where $\tilde{p}(x\vert y)$ is a conditional distribution defined
using a special time-reversed protocol (see Appendix~\ref{app:derivation}
for details). The formal similarity between (\ref{eq:dft}) and (\ref{eq:resInt})
is clear. However, unlike the DFT, our result makes no explicit reference
to a time-reversed process, and $-\log_{2}\tilde{p}(x\vert y)$ is
replaced by an algorithmic information term $\KKp$. Simply put, both
the DFT and our result show that the breaking of symmetry between
the forward and reverse maps must be paid for by heat generation.
However, the notion of symmetry breaking is defined differently in
these two results. Our result can be understood as a kind of ``algorithmic
fluctuation theorem'' which quantifies symmetry breaking in terms
of the \emph{algorithmic reversal}, rather than \emph{time reversal
}as in a regular DFT.}

\subsection{Generalized Zurek's bound}

\newstuff{We now derived a simplified version of our result, which
will add insight and highlight the connection to Zurek's bound (\ref{eq:zurek0}).
First, we combine the second inequality in (\ref{eq:df3}) with (\ref{eq:resInt2}),
while absorbing the additive constant into $\uuconst$, to give
\begin{equation}
\blnt\qqBase+\log_{2}\frac{1}{p(y\vert x)}+K(\PP)+\uuconst\ge K(x\vert y).\label{eq:resInt2Simpler}
\end{equation}
This bound, which is our second main result, separates those terms
that depend on the details of the protocol $\PP$ (left hand side)
from those terms that depends only on the logical computation $\xM$
(right hand side).} It implies that $\KK$ is an unavoidable algorithmic
cost of carrying out the computation $\xM$ with any protocol. This
cost must be paid with some combination of heat, noise, and protocol
complexity, implying a tradeoff between these three resources, which
is illustrated in \figref{1}. As we discuss in more detail below,
the inequality (\ref{eq:resInt2Simpler}) may be seen as a generalization
of Zurek's bound.

\newstuff{We emphasize that (\ref{eq:resInt2Simpler}) is generally
weaker than our first result (\ref{eq:resInt2}), because the inequality
$K(x\vert y)\le K(\PP)+\KKp+O(1)$ is not always tight. The difference
between these two bounds is illustrated in an example below. In general,
(\ref{eq:resInt2}) may be considered as the more fundamental and
achievable bound, while (\ref{eq:resInt2Simpler}) is a useful simplification
that allows us to separate physical from logical terms. In principle,
it is also possible to derive other bounds and decompositions starting
from (\ref{eq:resInt2}) and (\ref{eq:resInt2Simpler}). For example,
one could derive other bounds by decomposing the complexity term $K(\PP)$
in (\ref{eq:resInt2Simpler}) into contributions from different aspects
of the protocol (e.g., the complexity of the unitary versus the complexity
of the heat bath).

}

Finally, observe that the terms $\ppComp$ and $\uuconst$ in (\ref{eq:resInt2Simpler})
do not depend on $x$ and $y$. Thus, we may generally write
\[
\blnt\qqBase+\log_{2}\frac{1}{p(y\vert x)}+O(1)\ge K(x\vert y),
\]
where the additive constant $O(1)$ can now depend on the protocol.
In fact, even this additive constant may be disregarded in an appropriate
limit, leading to a simpler inequality. Suppose that the set of logical
states is countably infinite, as might represent the logical states
of some Turing machine. Then, consider repeating the same protocol
on a sequence of inputs $x_{1},x_{2},\dots$ of increasing length
($\ell(x_{n})=n$), producing a sequence of outputs $y_{1},y_{2},\dots$.
In the $n\to\infty$ limit, the heat per input bit can be bounded
without any additive constants as
\begin{equation}
\lim_{n\to\infty}\frac{\blnt Q(x_{n}\!\to y_{n})}{n}\ge\lim_{n\to\infty}\frac{K(x_{n}\vert y_{n})+\log_{2}p(y_{n}\vert x_{n})}{n},\label{eq:resIntLimit}
\end{equation}
assuming that the limits exist.

\subsection{Connection to measurable quantities}

How can one measure the terms that appear in our bounds, for instance
if one wishes to compare the theoretical predictions with empirical
observations? In general, the heat and noise terms can be estimated
using standard techniques, e.g., by running the process many times
starting from input $x$ and measuring energy transfer to the bath
and output $y$ in each run. The algorithmic information terms, such
as $K(x\vert y)$, $\KKp$ and $\ppComp$, present a bigger challenge.

In general, Kolmogorov complexity terms are uncomputable. However,
they can be upper bounded (to arbitrary tightness) by computable compression
algorithms \citep{livi08,baumeler2019free,zenil2020review}. For instance,
$\ppComp$ in (\ref{eq:resInt2Simpler}) can be upper bounded using
$n_{\PP}$, the number of bits needed to specify the effective parameters
that define the protocol $\PP=(V,U,\beta,\epsilon)$. Any such upper
bound on $\ppComp$ leads to a valid but weaker bound when plugged
into (\ref{eq:resInt2Simpler}). 

The terms $\KK$ and $\KKp$ can also be upper bounded using computable
compressions of $x$ with side information, for instance 
using ``Lempel-Ziv compression with side information'' or related schemes
\citep{ziv1984fixed,subrahmanya1995sliding,uyematsu2003conditional,cai2006algorithm}.
However, such computable estimates of $K(x\vert y)$ are upper (not
lower) bounds, and therefore they do not generally preserve the inequalities
when plugged into (\ref{eq:resInt2}) and (\ref{eq:resInt2Simpler}).
\newstuff{The estimation of $\KKp$ also faces the problem of finding
the minimal description of the protocol $\PP^{*}$. However, this
latter issue can be ignored for simple protocols with small $\ppComp$,
since in that case $\KKp\approx\KK$ from (\ref{eq:simp1}).}

Nonetheless, compression-based estimators are frequently used to approximate
Kolmogorov complexity in practice \citep{cilibrasi2005clustering,zenil2020review,kennel2004testing,dingle2018input,avinery2019universal,ro2022model}.
To the extent that these estimators are justified, our bounds can
be useful for bounding heat generation using estimates of $K(x\vert y)$
and/or $\log_{2}p(y\vert x)$. Furthermore, the problem of estimating
$\KK$ is avoided if (\ref{eq:resIntLimit}) is used to bound $K(x_{n}\vert y_{n})$
in terms of empirical measurements of $Q(x_{n}\!\to y_{n})$ and $\log_{2}p(y_{n}\vert x_{n})$
for large $n$.

\begin{figure*}[t]
\includegraphics[width=0.85\textwidth]{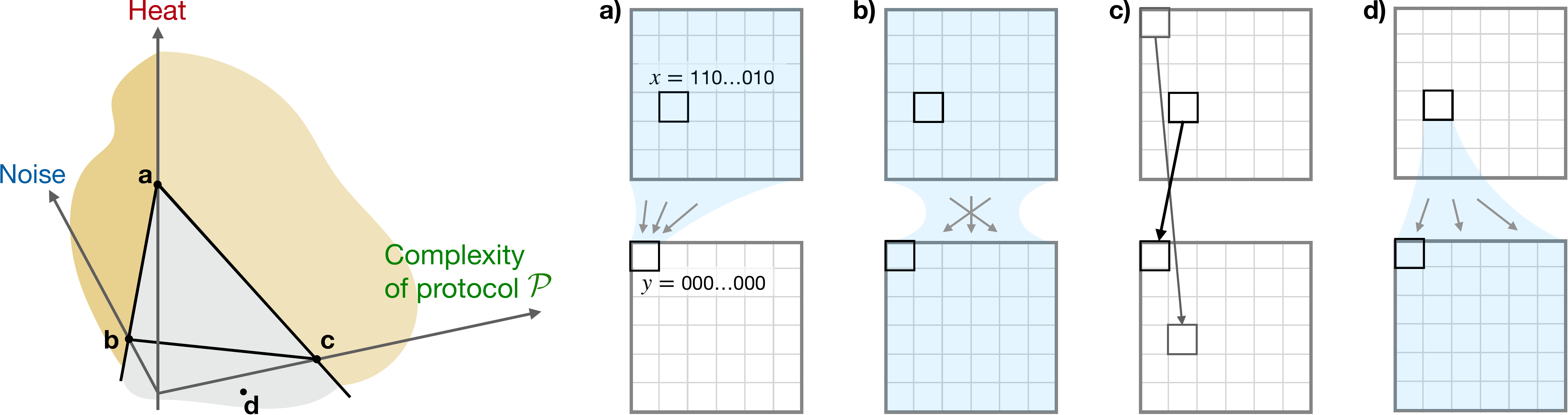}

\caption{\label{fig:2}Illustration of our results using the erasure task, where
an input string $x$ is mapped to an output $y$ consisting of all
0s, possibly in a noisy fashion. Four different processes that achieve
our bound (\ref{eq:resInt2Simpler}): \textbf{a)} Landauer erasure,
where all possible inputs are mapped to the target $y$. \textbf{b)}
Free expansion, where all inputs are allowed to thermalize to a uniform
equilibrium distribution. \textbf{c)} Swap, where a single input and
a single output are swapped, while all other states are left in place.
\textbf{d)} Controlled expansion, where a single input is allowed
to thermalize to a uniform equilibrium distribution.}
\end{figure*}

\section{Examples}

\label{sec:Example}

To make things concrete, we now illustrate our results using two examples.
In both examples, we consider a hard drive with $2^{n}$ logical states,
encoding all bit strings of length $n\gg1$. We consider two tasks.
The first is the ``erasure'' task mentioned in the Introduction,
which involves transforming a long incompressible string into a string
of all zeros. \newstuff{The second is a ``randomization'' task,
which involves transforming a string of all zeros into a long incompressible
output.}

\subsection{Example 1: Erasure}

We first consider the erasure task, where we map a long incompressible
input $x$ of length $n$ to an output $y$ consisting of $n$ zeros.
In this case, $K(x\vert y)\approx K(x)\approx n$. We sketch out four
different types of processes, illustrated in \figref{2}, that perform
the erasure. We demonstrate that each type of process saturates our
thermodynamic bounds (\ref{eq:resInt2}) and (\ref{eq:resInt2Simpler})
in a different way. In doing so, we also demonstrate that these two
bounds are equivalent for these processes.

\vspace{5pt}
\noindent 1) Landauer erasure, \figref{2}a). The Hamiltonian of the computational
subsystem $\soiX$ is set to zero and the subsystem is allowed to
relax to a uniform equilibrium. The Hamiltonian is then quasistatically
changed so that the system moves to an equilibrium distribution which
is (nearly) a delta-function centered at the all-0s string. In the
limit of quasistatic driving and peaked final energy functions, $\qqBase\approx\lntbIn n$
and $-\log_{2}p(y\vert x)\approx0$. This protocol has a simple description,
$\ppComp\approx0$, so (\ref{eq:simp1}) implies that our two bounds
(\ref{eq:resInt2}) and (\ref{eq:resInt2Simpler}) have the same minimal
cost, $\KKp\approx\KK\approx n$. Both bounds are achieved, with the
minimal cost paid  by heat.

\vspace{5pt}
\noindent 2) Free expansion, \figref{2}b). This is simply the first part of
the Landauer erasure process described above: the Hamiltonian of the
subsystem $\soiX$ is set to zero and the subsystem is again allowed
to relax to a uniform equilibrium. Since each final state is equally
likely under this free relaxation, $-\log_{2}p(y\vert x)=n$. No heat
is released or absorbed, $\qqBase=0$. This protocol has a simple
description, $\ppComp\approx0$, so our two bounds (\ref{eq:resInt2})
and (\ref{eq:resInt2Simpler}) again have the same minimal cost, $\KKp\approx\KK\approx n$.
Both bounds are achieved, with the minimal cost is paid for by noise.

\vspace{5pt}
\noindent 3) Swap, \figref{2}c). Subsystem $\soiX$ undergoes a unitary that
swaps the incompressible input string $x$ with the simple (all 0s)
output string $y$, while all other states are left untouched. This
process is deterministic, $-\log_{2}p(y\vert x)=0$, and no heat is
exchanged with the bath, $\qqBase=0$. However, this process requires
that the input state $x$ be ``hard coded'' into the the unitary
that implements the swap, so $K(\PP)\approx n$ and $\KKp\approx0$.
The minimal cost in (\ref{eq:resInt2}) vanishes, $\KKp\approx0$;
this bound is achieved since both heat and noise vanish. The minimal
cost in (\ref{eq:resInt2Simpler}) is $K(x\vert y)\approx n$; this
bound is again achieved and the cost is paid for by protocol complexity.

\vspace{5pt}
\noindent 4) Controlled expansion, \figref{2}d). The Hamiltonian of subsystem
$\soiX$ is initially set to a peaked function at the target input
$x$, and the subsystem is allowed to relax an equilibrium distribution
which is (nearly) a delta-function centered at $x$. The Hamiltonian
is then quasistatically changed to a constant energy function, so
that the system moves to a uniform equilibrium distribution. In the
limit of quasistatic driving and peaked initial energy function, heat
can be extracted from the bath, $\qqBase\approx-\lntbIn n$. However,
this process is noisy, $-\log_{2}p(y\vert x)\approx n$. In addition,
the initial Hamiltonian must include a description of the target input
state $x$. Therefore, as for the Swap, $K(\PP)\approx n$ and $\KKp\approx0$.
The minimal cost in (\ref{eq:resInt2}) vanishes, $\KKp\approx0$;
this bound is achieved since the heat and noise terms cancel. The
minimal cost in (\ref{eq:resInt2Simpler}) is $K(x\vert y)\approx n$;
this bound is also achieved and the cost is paid for by a combination
of absorbed heat, noise, and protocol complexity.

\newstuff{

\subsection{Example 2: Randomization}

\label{subsec:Example-2:-Randomization}

We now consider the randomization task, where an input $x$ consisting
of $n$ zeros is mapped to an output $y$ that is long and incompressible,
$K(y)\approx n$. Since the input is very simple, we have $K(x\vert y)\approx\KKp\approx0$.
From a logical perspective, this task may be considered to be the
opposite of erasure, in that the output and input roles are swapped.
However, as we will see, the thermodynamic properties of the two tasks
are very different.

We again consider the four types of processes mentioned in the previous
section. We show that for the randomization task, the two thermodynamics
bounds (\ref{eq:resInt2}) and (\ref{eq:resInt2Simpler}) can be different.
We also show that only some of the processes can saturate the tighter
bound (\ref{eq:resInt2}).

\vspace{5pt}
\noindent 1) Landauer reset. Subsystem $\soiX$ is allowed to freely relax to
a uniform equilibrium. The Hamiltonian is then quasistatically changed
so that the system moves to an equilibrium distribution which is (nearly)
a delta-function centered at the string $y$. In the limit of quasistatic
driving and peaked final energy functions, $\qqBase\approx\lntbIn n$
and $-\log_{2}p(y\vert x)\approx0$. Plugging into (\ref{eq:resInt2})
gives $n > 0$ (up to additive constants), so the bound is
not tight. Since the protocol must contain a description of $y$,
we also have $\ppComp\approx n$. Plugging into the weaker bound (\ref{eq:resInt2Simpler})
shows that it is even looser, $2n > 0$.

\vspace{5pt}
\noindent 2) Free expansion. Subsystem $A$ is allowed to relax to a uniform
equilibrium. Since each final state is equally likely under this free
relaxation, $-\log_{2}p(y\vert x)=n$, and no heat is released or
absorbed, $\qqBase=0$. Plugging into (\ref{eq:resInt2}) gives $n > 0$,
so the bound is not tight. In this case, the protocol is very simple,
$\ppComp\approx0$, so (\ref{eq:resInt2Simpler}) also gives $n > 0$.

\vspace{5pt}
\noindent 3) Swap. Subsystem $\soiX$ undergoes a unitary that swaps $x$ and
$y$, while all other states are left untouched. This process is deterministic,
$-\log_{2}p(y\vert x)=0$, and no heat is exchanged with the bath,
$\qqBase=0$. In this case, the bound (\ref{eq:resInt2}) is achieved.
However, the protocol must contain a description of $y$, so $\ppComp\approx n$.
Therefore, the weaker bound (\ref{eq:resInt2Simpler}) is not tight,
$n > 0$.

\vspace{5pt}
\noindent 4) Controlled expansion. The Hamiltonian of subsystem $\soiX$ is
initially set to a peaked function at the target input $x$, and the
system is allowed to relax to an equilibrium distribution which is
(nearly) a delta-function centered at $x$. The Hamiltonian is then
quasistatically changed to a constant energy function, so that the
system moves to a uniform equilibrium distribution. In the limit of
quasistatic driving and peaked initial energy function, heat can be
extracted from the bath, $\qqBase\approx-\lntbIn n$. However, this
process is noisy, $-\log_{2}p(y\vert x)\approx n$. The bound (\ref{eq:resInt2})
is achieved, since the noise and heat terms cancel. However, the protocol
must contain a description of $y$, so $\ppComp\approx n$. Therefore,
the weaker bound (\ref{eq:resInt2Simpler}) is not tight, $n > 0$.

\vspace{5pt}

As we see, although the erasure and randomization tasks are similar
from a logical standpoint, thermodynamically they are quite different.
In general, our bound (\ref{eq:resInt2}) can only be achieved when
the system is not allowed to thermalize to a uniform distribution
at the beginning of the process. This is because any such thermalization
dissipates initial order, i.e., it tends to map the simple input $x$
to a $n$-bit string sampled from the uniform distribution, which
almost always has high complexity \citep{livi08}. This is why erasure
and randomization tasks differ: in the erasure task, an initial thermalization
tends to replace the complex input $x$ with another complex string.
It may be said that the complex input in the erasure task is ``in
equilibrium'' with respect to the uniform distribution, while the
simple input in the randomization task is ``out of equilibrium''
with respect to the uniform distribution.}

\section{Relation to prior work}

\label{sec:Relation-to-prior}

We briefly compare our results with prior work.

We first consider Zurek's bound (\ref{eq:zurek0}), as proposed in
Ref.~\citep{zure89a}. That result is a special case of (\ref{eq:resInt2Simpler}),
which applies to deterministic computations ($-\log p(x\vert y)=0$).
Our bound is more general because it also applies to noisy processes,
and because it explicitly highlights the dependence on protocol complexity.
Perhaps more importantly, our bound is derived from a Hamiltonian
formulation, which extends it to both quantum and classical systems
while avoiding certain problematic aspects of the original derivation
(see also discussion in \citep{wolpert2019stochastic,kolchinsky2020thermodynamic}).

Let us review the derivation of the original result, which used a
two-step process \citep{zure89a}. First, the input $x$ is mapped
to the pair of strings $(y,s)$, where $s$ is an auxiliary binary
string that allows $x$ to be recovered from $(y,s)$ using some computer
program. By construction, this first step is logically reversible,
so in principle it can be carried out without generating heat. Second,
the string $s$, which is stored in $\ell(s)$ binary degrees of freedom, 
is reset to all 0s. The cost of this erasure was argued to be
$Q\ge\lntbIn\ell(s)$, as follows from Landauer's bound (\ref{eq:landauer0}).
Finally, (\ref{eq:zurek0}) follows since $x$ is decodable from $s$
and $y$, therefore $\ell(s)\ge K(x\vert y)+O(1)$ according to the
definition of conditional Kolmogorov complexity (\ref{eq:kdef}) (here
$O(1)$ is a constant that reflects the choice of the universal computer).

There are two problematic aspects of using Landauer's principle to
say that $\lntbIn\ell(s)$ is the minimal heat needed to reset $\ell(s)$
binary degrees of freedom (the same statement also appears in other
related work \citep{zure89a,bennett1993thermodynamics,bennett1998information,baumeler2019free}).
The first problem is that Landauer's bound constrains the average
heat across a statistical ensemble of inputs. For any individual input
$x$ and individual auxiliary string $s$, the reset protocol can
be designed to achieve much lower heat generation. In the extreme
case, $s$ may be ``hard coded'' into the reset protocol, so that
the reset generates no heat at all (as in the example shown in \figref{2}c).
For this reason, a proper accounting of the second step should also
include the complexity of the reset protocol, analogous to the term
$K(\PP)$ that appears in our result \footnote{It should be noted that the idea that the algorithmic complexity of
the ensemble should also be included in a proper thermodynamic accounting
appears in a different part of Zurek's paper, \citep[Eq.  (23), ][]{zure89a}.}. The second problem with the use of Landauer's bound is more technical.
Suppose that the reset protocol does not have any hard-coded information
about the string $s$. Then, in order to reset $\ell(s)$ binary degrees
of freedom at the Landauer cost of $\lntbIn\ell(s)$, the protocol
must have information about the string length $\ell(s)$, which will
vary between different inputs \citep{wolpert2019stochastic}. If this
length is measured before running the reset, then this information
about $\ell(s)$ also has to be erased, resulting in additional $O(\log\ell(s))$
heat generation. In fact, such logarithmic correction terms are unavoidable
when trying to erase an arbitrary binary string of arbitrary length.
(This problem can be avoided by appropriate use of self-delimiting
codes during the erasure process, 
as done in our construction in Appendix~\ref{app:Achievability}).

In our own recent work, we rederived a version of Zurek's bound (\ref{eq:zurek0})
using modern methods \citep{kolchinsky2020thermodynamic}. Although
our derivation did not make use of Landauer's bound as an intermediate
step, it only applied to classical and deterministic computations.
In addition, unlike the result presented here, it relied on idealized
assumptions (such as the assumption of an idealized bath).

Finally, our approach should be contrasted with ``single-shot thermodynamics''
as recently considered in quantum thermodynamics and quantum information
theory \citep{aaberg2013truly,faist2015minimal,rio2011thermodynamic,tomamichelQuantumInformationProcessing2016,horodecki2013fundamental}.
Single-shot thermodynamics focuses on the thermodynamic costs and
benefits incurred by individual computations at a guaranteed high
probability, relative to some input ensemble. For this reason, single-shot
costs are still defined in terms of statistical ensembles \citep{faist2015minimal}.
In general, the goals of single-shot thermodynamics are different
from our goals, which is to derive an algorithmic bound on the cost
of a (deterministic or noisy) computation that does not reference any
input ensemble.

\section{Discussion}

\label{sec:conclusion}

In this paper, we identified a fundamental thermodynamic bound on
the cost of an individual computation. Our bound, which makes no reference
to the statistical ensemble of inputs, was derived by combining ideas
from two different lines of research. The first is algorithmic information
theory (AIT), which defines information at the level of individual
strings, rather than statistical ensembles. The second is stochastic
thermodynamics, which defines thermodynamic quantities at the level
of individual trajectories, rather than ensembles of trajectories. 

The main assumption used to derive our results, and the most unusual
one in the physics literature, is the \emph{computability assumption},
which says that there exists a program that approximates the values
of $p(y,\bathf\vert x,\bathi)$, $\beta$, and $\epsilon_{b}$ to
arbitrary precision. It may be difficult to imagine a physical unitary
or observable which cannot be calculated numerically using a finite
program. Nonetheless, such non-computable unitaries and observables
do exist in a mathematical sense \citep{nielsen1997computable}, although
it is not clear whether they can be realized in any real-world physical
system. The so-called ``Physical Church-Turing thesis'' (PCT), whose
validity is the subject of ongoing discussion, postulates that non-computable
unitaries and observables are not physically realizable \citep{gandy1980church,deutsch1985quantum,geroch1986computability,braverman2015space}.
If the PCT is true, then our computability assumption is satisfied
by all protocols that can be realized in real-world physical systems.
If the PCT is not true, then there exists some physically realizable
protocol $\PP$ that does not satisfy the computability assumption
--- that is, it does not have a minimal description $\PP^{*}$, so
(\ref{eq:resInt2}) does not apply.  In general, our results point
to an interesting relationship between PCT and thermodynamics.

There are several possible directions for future research. As mentioned
in Section~\ref{subsec:Algorithmic-costs-of}, our results may be
understood as ``algorithmic fluctuation theorems'', that is algorithmic
versions of detailed fluctuation theorems from stochastic thermodynamics.
One may investigate there other results from stochastic thermodynamics,
such as thermodynamic uncertainty relations \citep{horowitz2020thermodynamic}
and thermodynamic speed limits \citep{shiraishi_speed_2018}, may
also have algorithmic analogues.

\newstuff{Another interesting research direction would extend our
approach to fully quantum computations. As discussed in Section~\ref{subsec:quantum},
even though our analysis applies to quantum processes, the computation
is operationalized using a two-point measurement scheme, in which
a fixed reference basis is used to initialize the input and determine
the output via a projective measurement. Therefore, the overall computation
that maps the input string $x$ to the output string $y$ is classical,
even if the process may use quantum effects at intermediate stages.
In future work, it may be interesting to investigate the thermodynamic
cost of a quantum computation that maps some input pure state $\vert\psi\rangle\langle\psi\vert$
to some output pure state $\vert\phi\rangle\langle\phi\vert$, without
assuming that $\vert\psi\rangle\langle\psi\vert$ and $\vert\phi\rangle\langle\phi\vert$
belong to any fixed basis or result from projective measurements.}

\section*{Acknowledgements}

We thank David Wolpert, Hanzhi Jiang, and members of the Sosuke Ito
lab for helpful discussions, and the Santa Fe Institute for helping
to support this research. This research was partly supported by Foundational
Questions Institute (FQXi) Grant Number FQXi-RFP-IPW-1912 and JSPS
KAKENHI Grant Number JP19H05796.

\appendix
\renewcommand{\appendixname}{APPENDIX}

\section{DERIVATION OF (\ref{eq:resInt2})}

\label{app:derivation}

Here we derive our main result, (\ref{eq:resInt2}). We begin by proving
the following useful inequality,
\begin{equation}
-\log_{2}p(y\vert x)+\qq\ge-\log_{2}\revP(x\vert y),\label{eq:dft-1}
\end{equation}
where $p(y\vert x)$ is defined as in (\ref{eq:cond1}) and
\begin{equation}
\revP(x\vert y):=\sum_{\bathi,\bathf}p(y,\bathf\vert x,\bathi)\pi(\bathf).\label{eq:qdef}
\end{equation}
Note that $p(y,\bathf\vert x,\bathi)=\vert\langle y,\bathf\vert U\vert x,\bathi\rangle\vert^{2}=\vert\langle x,b\vert U^{\dagger}\vert y,c\rangle\vert^{2}$, which 
can be interpreted as the conditional probability of observing final
state $\vert x,b\rangle$ given initial state $\vert y,c\rangle$
under the adjoint unitary evolution $U^{\dagger}$. Therefore $\sum_{x,b}p(y,\bathf\vert x,\bathi)=1$
and $\sum_{\xs}\revP(x\vert y)=1$, meaning that $\revP(x\vert y)$
is a normalized conditional probability distribution. In addition,
the adjoint unitary $U^{\dagger}$ can be understood as the time-reversal
of the actual unitary $U$ \citep{campisi2011colloquium,manzanoQuantumFluctuationTheorems2018}.
Therefore, $\revP(x\vert y)$ can also be seen as the probability
of output $x$ from input $y$ under the time-reversed process.

\newstuff{Next, write the following identity
\begin{equation}
p(y,\bathf\vert x,\bathi)\pi(\bathf)=p(y,\bathf\vert x,\bathi)\pi(b)e^{-\beta(\nrg_{\bathf}-\nrg_{\bathi})}.\label{eq:tiled2}
\end{equation}
We sum both sides of (\ref{eq:tiled2}) over the initial and final
bath states $\bathi$ and $\bathf$ to give
\begin{align}
\revP(x\vert y) & =\sum_{\bathi,\bathf}p(y,\bathf\vert x,\bathi)\pi(b)e^{-\beta(\nrg_{\bf}-\nrg_{\bs})}\nonumber \\
 & =p(y\vert x)\sum_{\bathi,\bathf}\frac{p(y,\bathf\vert x,\bathi)\pi(\bathi)}{p(y\vert x)}e^{-\beta(\nrg_{\bf}-\nrg_{\bs})}\nonumber \\
 & \ge p(y\vert x)e^{-\beta\qqBase}.\label{eq:res2}
\end{align}
In the first line, we used the definition of $\revP(x\vert y)$ from
(\ref{eq:qdef}). In the second line, we used the definition of $p(y\vert x)$
from (\ref{eq:cond1}). In the last line, we used Jensen's inequality
and the definition of $\qqBase$ from (\ref{eq:heat-def}). Rearranging
(\ref{eq:res2}) gives (\ref{eq:dft-1}).} The inequality (\ref{eq:res2})
becomes tight when heat fluctuations become small, which is expected
in the limit of a large self-averaging heat bath and slow driving.
This inequality appears as (\ref{eq:dft}) in the main text. We note that this
type of derivation is often used to prove detailed fluctuation theorems
in stochastic thermodynamics \citep{jarzynski2000hamiltonian,sagawa2012fluctuation,kwonFluctuationTheoremsQuantum2019,BuscemiBayesianRetrodiction2021,jarzynski2004classical,manzanoQuantumFluctuationTheorems2018}. 

\newstuff{Next, we introduce the bound
\begin{equation}
-\log_{2}\revP(x\vert y)+\gamma_{\utm}\ge\KKp,\label{eq:dd}
\end{equation}
where $K(x\vert y,\PP^{*})$ is the conditional Kolmogorov complexity
described in Section~\ref{subsec:Computability-assumption}, and 
$\gamma_{\utm}$ is a constant that does not depend on $x$, $y$,
or $\PP$. While (\ref{eq:dd}) is a standard result in AIT \citep{vitanyi2013conditional},
here we sketch out its proof at a high level. Recall that $\PP^{*}$
is a program that computes the values of $p(y,\bathf\vert x,\bathi)$
and $\pi(\bathi)$ to arbitrary precision (\ref{eq:PPprog}). Next,
we define the following program which outputs $x$ when provided with
$y$ and $\PP^{*}$ as input:
\begin{enumerate}
\item It calculates the conditional probability distribution $x'\mapsto\revP(x'\vert y)$
for the given $y$ and all $x'\in\{0,1\}^{*}$. This is done by running
the provided program $\PP^{*}$ and computing $\revP(x'\vert y)$
via (\ref{eq:qdef}).
\item Using the conditional probability distribution $\revP(x'\vert y)$,
it constructs a \emph{codebook}, a function that maps codewords $s\in\{0,1\}^{*}$
to strings $x'\in\{0,1\}^{*}$. Using a standard coding algorithm,
such as Shannon-Fano coding or Huffman coding, the codeword $s_{x'}$
assigned to string $x'$ may be chosen to have length $\ell(s_{x'})\le\left\lceil -\log_{2}\revP(x'\vert y)\right\rceil $.
\item The program contains a copy of the codeword $s_{x}$ that maps to
string $x$. It looks up this codeword in the generated codebook and
prints the resulting string $x$.
\end{enumerate}
The length of this program is no more than 
\[
\ell(s_{x})+\gamma_{\utm}\le-\log_{2}\revP(x'\vert y)+\gamma_{\utm}+1,
\]
where $\gamma_{\utm}$ is an additive constant that reflect the length
of the algorithm needed to compute the conditional distribution $\revP(x'\vert y)$
(while calling $\PP^{*}$ as a subroutine), generate the codebook,
and look up the codeword $s_{x}$. The bound (\ref{eq:dd}) follows after
absorbing 1 into the additive constant $\gamma_{\utm}$, since $\KKp$
is the length of the shortest program that outputs $x$ when provided
with $y$ and $\PP^{*}$ as input.}

Combining with (\ref{eq:dd}) and (\ref{eq:dft-1}) gives (\ref{eq:resInt2}).

\section{ACHIEVABILITY OF (\ref{eq:resInt2})}

\label{app:Achievability}

\begin{table*}
\begin{tabular*}{1\textwidth}{@{\extracolsep{\fill}}|>{\centering}p{0.08\textwidth}|>{\raggedright}p{0.5\textwidth}|>{\centering}p{0.2\textwidth}|>{\centering}p{0.15\textwidth}|}
\hline 
\textbf{Step} & \textbf{Description} & \textbf{State of $X\times M$ at end of step} & \textbf{Heat for input $x$ and output $y$}\tabularnewline
\hline 
\hline 
\emph{Initialize} & Subsystems $X\times M$ begin with the initial Hamiltonian
\begin{equation}
H_{XM}=\sum_{x',m}(\chi_{x'}+\mu_{m})\vert x',m\rangle\langle x',m\vert.\label{eq:h1}
\end{equation}
 & $\vert x,\varnothing\rangle\langle x,\varnothing\vert$ & \tabularnewline
\hline 
\emph{Copy} & Input state $\vert x\rangle$ is copied from $X$ into $M$ using
unitary over $X\times M$. & $\vert x,x\rangle\langle x,x\vert$ & \tabularnewline
\hline 
\emph{Quench 1} & Hamiltonian $H_{XM}$ is quenched from (\ref{eq:h1}) to
\begin{equation}
H_{XM}=\sum_{x',m}(\chi_{x'}+\mu_{m}-\eta\delta_{x',m})\vert x',m\rangle\langle x',m\vert.\label{eq:h2}
\end{equation}
 & $\vert x,x\rangle\langle x,x\vert$ & \tabularnewline
\hline 
\emph{Relax 1} & $X$ freely relaxes to conditional equilibrium for $H_{XM}$, while
$M$ is held fixed. & $\approx\vert x,x\rangle\langle x,x\vert$ & $\approx0$\tabularnewline
\hline 
\emph{Compute} & Hamiltonian $H_{XM}$ is quasistatically changed to
\begin{equation}
H_{XM}=\sum_{y',m}[-\beta^{-1}\ln p(y'\vert m)+\mu_{m}]\vert y',m\rangle\langle y',m\vert,\label{eq:h3}
\end{equation}
during which $X$ is allowed to relax to equilibrium and $M$ is held
fixed. & $\approx{\displaystyle \sum_{y'}p(y'\vert x)\vert y',x\rangle\langle y',x\vert}$ & $\approx\beta^{-1}\ln p(y\vert x)$\tabularnewline
\hline 
\emph{Quench 2} & Hamiltonian $H_{XM}$ is quenched from (\ref{eq:h3}) to
\begin{equation}
H_{XM}=\sum_{y',m}\omega_{y'm}\vert y',m\rangle\langle y',m\vert.\label{eq:h4}
\end{equation}
 &  & \tabularnewline
\hline 
\emph{Relax 2} & $M$ freely relaxes to conditional equilibrium for (\ref{eq:h4})
while $X$ is held fixed. & $\approx{\displaystyle \sum_{y',m}p(y'\vert x)q_{\omega}^{\text{eq}}(m\vert y')}$
$\times\vert y',m\rangle\langle y',m\vert$ & $\approx\omega_{yx}-\langle\omega\rangle_{q_{\omega}^{\text{eq}}(M\vert y)}$\tabularnewline
\hline 
\emph{Reset} & Hamiltonian $H_{XM}$ is quasistatically changed to 
\begin{equation}
H_{XM}=\sum_{y',m}(\chi_{y'}+\mu_{m}-\eta\delta_{\varnothing,m})\vert y',m\rangle\langle y',m\vert,\label{eq:h5}
\end{equation}
during which $M$ is allowed to relax to equilibrium and $X$ is held
fixed. & $\approx{\displaystyle \sum_{y'}p(y'\vert x)\vert y',\varnothing\rangle\langle y',\varnothing\vert}$ & $\approx\text{\ensuremath{\beta^{-1}}}S(q_{\omega}^{\text{eq}}(M\vert y))$\tabularnewline
\hline 
\emph{Final quench} & Hamiltonian $H_{XM}$ is quenched to the initial Hamiltonian in (\ref{eq:h1}). & $\approx{\displaystyle \sum_{y'}p(y'\vert x)\vert y',\varnothing\rangle\langle y',\varnothing\vert}$ & \tabularnewline
\hline 
\end{tabular*}

\caption{\label{tab:achieve}Description of the 9-step protocol that implements
a stochastic input-output map that is arbitrarily close to some desired
$p(y\vert x)$, while achieving a heat dissipation arbitrarily close
to the bound (\ref{eq:resInt2}). $\delta$ is the Kronecker delta
function.}
\end{table*}

Here we show that the bound (\ref{eq:resInt2}) can be achieved. 

\subsection{Protocol Construction}

Formally, we imagine that we are provided with some desired input-output
conditional probability distribution $p_{Y\vert X}$. We also imagine
that we are provided with a thermal environment with 
inverse temperature $\beta$ and energy function $\epsilon$. We assume
that these are all computable, meaning that there exists some computer
program that can output the value of any $p(y\vert x)$, any $\epsilon_{b}$,
and $\beta$ to any desired degree of precision. We also assume that
the heat bath described by $\epsilon$ and $\beta$ is nearly ideal,
i.e., it is large and undergoes rapid self-equilibration. Then, in
the limit of an ideal bath, we show that there exists a protocol $\PP$
that can come arbitrarily closely to the bound (\ref{eq:resInt2}). We
do so by sketching out the construction of such a protocol $\PP$
at a high level, without delving into full technical rigor which would
go beyond the scope of this paper.

Our construction will suppose that subsystem $X$ has access to an
additional ``memory subsystem'' $M$. The memory subsystem has the
same dimensionality as $X$ and it acts as a catalyst: it is initialized
in an unentangled ``empty string'' pure state $\vert\varnothing\rangle$
at the beginning of the protocol, and it is left in (nearly) the same
pure state after the protocol finishes. We will perform protocols
on $X$ that depend on the state of $M$ while holding the state of
$M$ fixed --- and vice versa for protocols over $M$ with $X$ held
fixed. Such protocols are often termed ``feedback control'' in the
literature \citep{sagawa2008second,parrondo2015thermodynamics}.

We will also use the fact that a nearly ideal bath can be weakly coupled
to subsystems $X\times M,$ and then used to carry out transformations
over $X\times M$ in a (nearly) quasistatic and thermodynamically
reversible manner (see constructions in \citep{anders2013thermodynamics,reeb_improved_2014,aaberg2013truly,skrzypczyk2014work}).
Weak coupling also allows us to write the First Law as $Q=W-\Delta E_{XM}$,
where $Q$ is generated heat, $W$ is work done on the system, and
$\Delta E_{XM}$ is the expected energy increase of subsystems $X\times M$.

Our protocol consists of 9 steps, which are described in \tabref{achieve}.
The third column in \tabref{achieve} shows the approximate state
of subsystems $X\times M$ at the end of each step. (The approximations
become exact in the limit of slow driving, idealized bath, complete
relaxation to equilibrium, and infinitely peaked energy functions
$\eta\to\infty$.) It can be verified that the protocol maps $X$
from input state $\vert x\rangle$ to output state $\vert y\rangle$
sampled from $\approx p(y\vert x)$, while the memory subsystem $M$
starts and ends arbitrarily close to the pure state $\vert\varnothing\rangle$.
Note also that the protocol starts and ends on the same Hamiltonian,
(\ref{eq:h1}). The protocol is effectively classical, in that it
does not exploit coherence or entanglement, even though it may be
implemented on a quantum system.

We emphasize a few aspects of our construction. First, the \emph{Copy
}step does not violate the no-cloning theorem, since $\vert x\rangle$
is assumed to come from the fixed orthonormal basis $\{\vert\sysstate\rangle\}_{\sysstate\in\{0,1\}^{*}}$.
Second, two of the steps \emph{(Compute} and \emph{Reset}) involve
thermodynamically reversible feedback control operations, as discussed
above (see also \citep[Section 5.1, ][]{reeb_improved_2014}). Third,
there are three energy functions $\chi,\mu,\omega$ and a scalar $\eta$
that appear in our construction, which may be chosen somewhat arbitrarily.
The energy functions $\chi$ and $\mu$, which appear in (\ref{eq:h1}),
(\ref{eq:h2}), (\ref{eq:h3}), and (\ref{eq:h5}), refer to arbitrary
``baseline'' energy values for $X$ and $M$ respectively. The scalar
$\eta\gg0$ refers to a large energy value that favors equilibrium
correlations between $X$ and $M$ under the Hamiltonian (\ref{eq:h2}),
and favors equilibrium reset of $M$ to state $\vert\varnothing\rangle$
under the Hamiltonian (\ref{eq:h5}). We usually consider very large
$\eta$, approaching the limit $\eta\to\infty$. The energy function
$\omega$ in (\ref{eq:h4}) refers to a set of energy values over
$X\times M$, which we will return to below. Below we will assume
that $\chi,\mu,\omega$ and $\eta$ are algorithmically simple (i.e.,
there exists a short program to compute their values). 

We now calculate the heat generated by this protocol during the computation
$\xM$. The approximate amount of heat generated by each step of the
protocol, conditioned on input $\vert x\rangle$ and output $\vert y\rangle$,
is shown in the last column of \tabref{achieve} (as above, the approximations
become exact under appropriate limits). We describe the calculated
heat values in more depth. \emph{Copy} involves a unitary over $X\times M$,
so it does not involve any heat exchange. The three quench steps (\emph{Quench}
\emph{1, Quench 2, Final Quench}) refer to (nearly) instantaneous
changes of the Hamiltonian, so they also do not involve any heat exchange.

\emph{Relax 1} does not involve driving, therefore the heat is given
by the decrease of the energy of subsystem $X\times M$ due to the
change of the statistical state. In fact, the statistical state does
not change (in the limit $\eta\to\infty$) and so heat vanishes. For\emph{
Relax 2}, the heat is also equal to the decrease of the energy of
$X\times M$ due to the change of the statistical state. Conditioned
on input $x$ and output $y$, subsystems $X\times M$ are (nearly)
in pure state $\vert y,x\rangle\langle y,x\vert$ at the beginning
of this step. At the end of this step, subsystem $M$ is found in
the conditional equilibrium distribution 
\begin{equation}
q_{\omega}^{\text{eq}}(M=m\vert X=y)=e^{-\beta(\omega_{ym}-F_{M\vert X=y}^{\omega})},\label{eq:gf}
\end{equation}
where we defined the conditional free energy
\begin{equation}
F_{M\vert X=y}^{\omega}=-\frac{1}{\beta}\ln\sum_{m}e^{-\beta\omega_{ym}}.\label{eq:fe2}
\end{equation}

To calculate the heat generated by \emph{Compute}, we use the First
Law of Thermodynamics, $Q=W-\Delta E_{XM}$. We also use that in the
quasistatic and classical limit, work is equal to the increase of
the equilibrium free energy of $X\times M$ and work fluctuations
vanish \citep{speck2004distribution,scandi2020quantum}. Since $M$
is held fixed in state $\vert x\rangle$, we consider the increase
of the conditional free energy, $W\approx\Delta F_{X\vert M=x}$.
Then, observe that in this step, $X\times M$ is transformed from
the pure state $\vert x,x\rangle\langle x,x\vert$ with Hamiltonian
(\ref{eq:h2}) to the mixed state $\sum_{y'}p(y'\vert x)\vert y',x\rangle\langle y',x\vert$
with Hamiltonian (\ref{eq:h3}), so 
\begin{align*}
\Delta F_{X\vert M=x} & =-\frac{1}{\beta}\ln\sum_{y}e^{\ln p(y\vert x)-\beta\mu_{x}}-(\chi_{x}+\mu_{x}-\eta)\\
 & =\mu_{x}-(\chi_{x}+\mu_{x}-\eta).
\end{align*}
The change of energy of $X\times M$, conditioned on input $x$ and
output $y$, is $\Delta E_{XM}\approx-\beta^{-1}\ln p(y\vert x)+\mu_{x}-(\chi_{x}+\mu_{x}-\eta)$.
Plugging into $Q\approx\Delta F_{X\vert M=x}-\Delta E_{XM}$ gives
$Q\approx\beta^{-1}\ln p(y\vert x)$, as found in \tabref{achieve}.

For \emph{Reset}, we use that the heat generated by a thermodynamically
reversible feedback control operation, where $M$ is modified while
$X$ is held fixed, is $\beta^{-1}$ times the decrease of the conditional
Shannon entropy $S(M\vert X)$ \citep{reeb_improved_2014}. Since
subsystem $X$ is observed in output state $y$ at the end of the
process, and since it does not change in between \emph{Reset} and
the end of the process, it must be in state $y$ during this step.
Thus, we consider the decrease of the conditional Shannon entropy
$S(M\vert X=y)$. The conditional distribution over energy eigenstates
of $M$ at the beginning of this step is (approximately)
$q_{\omega}^{\text{eq}}(M=m\vert X=y)$. At the end of this step,
$M$ is (nearly) in the pure state $\vert\varnothing\rangle$, independently
of the state of $X$, so the conditional Shannon entropy vanishes.
Combining gives the result in \tabref{achieve}.

Summing together the heat values in \tabref{achieve} implies that
the overall generated heat is 
\begin{equation}
\qqBase\approx\frac{1}{\beta}\ln p(y\vert x)+\omega_{yx}-F_{M\vert X=y}^{\omega},\label{eq:qq1}
\end{equation}
where we used the identity $F_{M\vert X=y}^{\omega}=\langle\omega\rangle_{q_{\omega}^{\text{eq}}(M\vert y)}-\beta^{-1}S(q_{\omega}^{\text{eq}}(M\vert y))$.

\global\long\def\RR{\mathcal{R}}%

\newstuff{

\subsection{Sketch of proof of achievability}

We now show that the protocol $\PP$ described above can approach the
bound (\ref{eq:resInt2}). 
We imagine there is some desired computation $p_{Y|X}$, as well as a nearly ideal heat bath described by the inverse temperature $\beta$ and energy function $\epsilon$. We  use $\mathcal{R}=(p_{Y|X}, \epsilon, \beta)$ to refer to the details of the computation. Importantly, we assume that $\mathcal{R}$ is computable. We show that the corresponding nine-step protocol $\mathcal{P}$, as defined in the previous subsection, comes close to equality in (\ref{eq:resInt2}). 
 
It must be emphasized that in the following, the protocol $\mathcal{P}$ is treated as a function of the desired computation $\mathcal{R}$. For this reason, the additive constant $O(1)$ that appears below refers to a quantity that doesn't depend on $x$, $y$, or the desired computation $\mathcal{R}$.  Similarly, within the family of protocols defined in the last subsection, $\gamma_\mathfrak{C}$ in (\ref{eq:resInt2}) should be understood as an additive constant that doesn't depend on $x$, $y$, or the details of the computation $\mathcal{R}$.

To begin, we define some algorithmic information measures. Let $\mathcal{R}^{*}\in\{0,1\}^{*}$
refer to the shortest program for the universal computer $\utm$ that
computes the conditional distribution $p_{Y\vert X}$, energy function
$\epsilon$, and inverse temperature $\beta$ to arbitrary precision.
In addition, let $K(x\vert y,\mathcal{R}^{*})$ indicate the length of the shortest
program that produces $x$ when provided with $y$ and $\mathcal{R}^{*}$ as
side information and let $K(\RR\vert y,\PP^{*})$ indicate the length
of the shortest program for computing $(p_{Y\vert X},\epsilon,\beta)$
given $y$ and $\PP^{*}$ as side information (recall that $\PP^{*}$
is the shortest program for computing the values (\ref{eq:PPprog})
for our protocol). Finally, let $K(\PP\vert y,\mathcal{R}^{*})$ indicate the
length of the shortest program for computing values (\ref{eq:PPprog})
given $y$ and $\mathcal{R}^{*}$ as side information.

In addition, let $\mathfrak{D}$ indicate some computable compression
algorithm with side information (e.g., Lempel-Ziv compression with
side information). Let $\ell_{\mathfrak{D}}$ indicate the length
of a program implementing this compression algorithm. 
Importantly, we assume that this algorithm is simple in the sense that 
$\ell_{\mathfrak{D}}=O(1)$. In addition, let
$c(x\vert y,\mathcal{R}^{*})$ indicate the length of the self-delimiting codeword produced
by this algorithm for $x$ when provided with side information $y$
and $\mathcal{R}^{*}$. Without loss of generality, we assume that the codebook
that defines $c(x\vert y,\mathcal{R}^{*})$ is complete for all $y$, so that Kraft's inequality
holds with equality \citep{cover_elements_2006,livi08}, $\sum_{x}2^{-c(x\vert y,\mathcal{R}^{*})}=1$.

Recall that we are free to choose any set of energy values $\omega$
in (\ref{eq:h4}), as long as the conditional free energy $F_{M\vert X=y}^{\omega}$
is finite and $q_{\omega}^{\text{eq}}(x\vert y)$ is a normalized
conditional distribution. We define these energy values $\omega$
as
\begin{equation}
\blnt\omega_{yx}=\ell_{\mathfrak{D}}+c(x\vert y,\mathcal{R}^{*}),\label{eq:ddd}
\end{equation}
Using $\sum_{x}2^{-c(x\vert y,\mathcal{R}^{*})}=1$ and (\ref{eq:fe2}), we calculate
that $F_{M\vert X=y}^{\omega}=\frac{\ln2}{\beta}\ell_{\mathfrak{D}}$.
Plugging into (\ref{eq:qq1}) and rearranging gives
\begin{equation}
\blnt\qqBase\approx\log_{2}p(y\vert x)+c(x\vert y,\mathcal{R}^{*}).\label{eq:g3}
\end{equation}
Note that better compression algorithms, which have smaller code lengths,
lead to lower heat production.

A known result from AIT states that conditional Kolmogorov complexities
obey a type of ``triangle inequality'' \citep[Lemma 3.9.1, ][]{livi08},
\begin{align*}
K(x\vert y,\PP^{*}) & \le K(\RR\vert y,\PP^{*})+K(x\vert y,\mathcal{R}^{*})+O(1)\\
K(x\vert y,\mathcal{R}^{*}) & \le K(\PP\vert y,\mathcal{R}^{*})+K(x\vert y,\PP^{*})+O(1).
\end{align*}
We will use this to show that 
\begin{equation}
K(x\vert y,\mathcal{R}^{*})=K(x\vert y,\PP^{*})+O(1).\label{eq:equiv3}
\end{equation}
Suppose that one is given $\mathcal{R}^{*}$ (which computes values of $p_{Y\vert X}$,
$\epsilon$, and $\beta$) along with a program that computes the
values of $\chi,\mu,\eta$ and $\omega$ that appear in the 9-step
protocol outlined in Table~\ref{tab:achieve}. In that case, one
could compute not only $\epsilon$ and $\beta$, but also the unitary
$U$ that implements that protocol $\PP$, giving the values of $p(y,\bathf\vert x,\bathi)=\vert\langle y,\bathf\vert U\vert x,\bathi\rangle\vert^{2}$
in (\ref{eq:PPprog}). Thus, we may write
\[
K(\PP\vert y,\mathcal{R}^{*})\le K(\chi,\mu,\eta\vert y,\mathcal{R}^{*})+K(\omega\vert y,\mathcal{R}^{*}) + O(1).
\]
Since $\chi$, $\mu$ and $\eta$ are arbitrary, we may assume that
they can be produced by a simple program when provided with $y$ and
$\mathcal{R}^{*}$ as side information, so $K(\chi,\mu,\eta\vert y,\mathcal{R}^{*})=O(1)$.
(Note that the scaling $\eta\to\infty$ can be achieved without additional
algorithmic cost by scaling $\eta$ as a function of the heat bath energy function
$\epsilon$, as specified by $\mathcal{R}^{*}$). The values of $\omega$
are determined by $\beta$ plus the choice of the compression algorithm
$\mathfrak{D}$, which obeys $\ell(\mathfrak{D})=O(1)$, hence $K(\omega\vert y,\mathcal{R}^{*})=O(1)$.
Combining implies that $K(x\vert y,\mathcal{R}^{*})\le K(x\vert y,\PP^{*})+O(1)$.
But it is also the case that $K(\RR\vert y,\PP^{*})=O(1)$, since
if one had a program to compute the values (\ref{eq:PPprog}) that
define $\PP$, one could easily compute $\epsilon$, $\beta$, and
$p_{Y\vert X}$ (the latter via (\ref{eq:cond1})). Therefore $K(x\vert y,\PP^{*})\le K(x\vert y,\mathcal{R}^{*})+O(1)$.
Combining implies (\ref{eq:equiv3}).

Finally, we have that
\begin{align}
c(x\vert y,\mathcal{R}^{*}) &= \ell_{\mathfrak{D}}+c(x\vert y,\mathcal{R}^{*})+O(1) \nonumber\\
& \ge K(x\vert y,\mathcal{R}^{*})+O(1)\label{eq:kk0}\\
 & =K(x\vert y,\PP^{*})+O(1),\label{eq:kk}
\end{align}
where in the first line we used $\ell_{\mathfrak{D}}=O(1)$, in the second line the definition of $K(x\vert y,\mathcal{R}^{*})$, and in the last line (\ref{eq:equiv3}).
Combining (\ref{eq:g3}) and (\ref{eq:kk}) then implies the bound
\begin{align}
\frac{\beta}{\ln 2}Q(x\shortrightarrow y)  & \ge
    \log_{2}p(y\vert x)+K(x\vert y,\mathcal{P}^{*})+O(1).\label{eq:gbf3}
\end{align}

The inequality in (\ref{eq:gbf3}) can be made arbitrarily
tight under an appropriate limit, which implies the achievability
of our bound (\ref{eq:resInt2}). Specifically, the inequality (\ref{eq:g3})
can be made arbitrarily tight in the limit of slow driving, idealized
bath, complete relaxation to equilibrium, and infinitely peaked energy
functions $\eta\to\infty$. The inequality (\ref{eq:kk0}) can be
made arbitrarily tight in the limit of increasingly good compression
algorithms, so that $c(x\vert y,\mathcal{R}^{*})$ approaches the ultimate algorithmic bound
$K(x\vert y,\mathcal{R}^{*})$ in (\ref{eq:kk0}). In fact, there are known techniques
to construct a sequence of compression algorithms such that $c(x\vert y,\mathcal{R}^{*})$
approaches $K(x\vert y,\mathcal{R}^{*})$ for any finite set of $x$ and $y$.
Such techniques are sometimes called ``dovetailing'' algorithms
in the literature \citep{livi08}. (We note that the compression algorithm
$\mathfrak{D}$ can ``scale'' the dovetailing parameter in line
with the scaling of the heat bath, as specified by the side information
$\RR^{*}$, without incurring an additional algorithmic cost.)

At the same time, Kolmogorov complexity is uncomputable, so the
convergence $c(x\vert y,\mathcal{R}^{*})\to K(x\vert y,\mathcal{R}^{*})$ cannot be uniform across $x$
and $y$, and in general one cannot know how far away from the ultimate
limit is any given compression algorithm \citep{livi08}. To summarize,
our bound (\ref{eq:resInt2}) is achievable in the sense that it is
possible to construct a sequence of thermal environments and physical
protocols that come arbitrarily close to equality for any finite set
of $x$ and $y$. At the same time, it is impossible to know how far
in the sequence one must go in order to guarantee that, for any desired
$x$ and $y$, the deficit in (\ref{eq:resInt2}) is bounded by a constant.

}

\bibliographystyle{IEEEtran}
\bibliography{main}

\clearpage
\end{document}